\documentstyle[epsfig]{lamuphys}
\makeatletter
\let\chapter\hid@chapter
\makeatother
\newcommand{\be}{\begin{equation}}
\newcommand{\ee}{\end{equation}}
\newcommand{\bea}{\begin{eqnarray}}
\newcommand{\eea}{\end{eqnarray}}
\newcommand{\lb}{\label}

\begin{document}
\pagenumbering{arabic}
\title{Decoherence: Concepts and Examples}

\author{Claus\,Kiefer\inst{1} and Erich\,Joos\inst{2}}

\institute{Fakult\"at f\"ur Physik, Universit\"at Freiburg,
Hermann-Herder-Stra\ss e 3, D-79104 Freiburg, Germany
\and
Rosenweg 2, D-22869 Schenefeld, Germany}

\maketitle

\begin{abstract}
We give a pedagogical introduction to the process of decoherence --
the irreversible emergence of classical properties through interaction
with the environment. After discussing the general concepts, we
present the following examples: Localisation of objects, quantum Zeno
effect,
classicality of fields and charges in QED, and decoherence in gravity
theory. We finally emphasise the important interpretational features
of decoherence.
\end{abstract}

\small
\noindent Report
     Freiburg THEP-98/4; to appear in {\em Quantum Future},
     edited by P. Blanchard and A. Jadczyk
     (Springer, Berlin, 1998).
\normalsize

\section{Introduction}

Since this conference is devoted to {\em Quantum Future}, i.e.,
to the future of research in fundamental (interpretational)
problems of quantum theory, it may be worthwile to start with a
brief look back to the {\em Quantum Past}. The Fifth Solvay Congress
in October~1927 marked both the completion of the {\em formal}
framework of quantum mechanics as well as the starting point of the
ongoing {\em interpretational} debate. The first point is clearly
expressed by Born and Heisenberg, who remarked at that congress
(Jammer~1974)
\begin{quote}
We maintain that quantum mechanics is a complete theory; its basic
physical
and mathematical hypotheses are not further susceptible of
modifications.
\end{quote}
The confidence expressed in this quotation has been confirmed by
the actual development: Although much progress has been made, of course,
in elaborating the formalism, particularly in quantum field theory,
its main elements, such as the superposition principle and the
probability interpretation as encoded in the Hilbert space formalism,
have been left unchanged. This is even true for tentative 
frameworks such as GUT theories or superstring theory. 
Although the latter may seem ``exotic" in some of its aspects
(containing
D-branes, many spacetime dimensions, etc.), it is very
traditional in the sense of the quantum theoretical formalism employed.

The starting point of the interpretational debate is marked
by the thorough discussions between Einstein and Bohr about the
meaning of the formalism. This debate was the core of most of the
later interpretational developments, including the EPR discussion,
the Bohm theory, and Bell's inequalities. That no general consensus
about the interpretation has been reached, is recognisable from
the vivid discussions during this conference.
Still, however, much progress has been made in the ``Quantum Past".
It has been clarified which questions can be settled by experiments
and which questions remain at present a matter of taste. 

Our contribution is devoted to one problem which plays a major role
in {\em all} conceptual discussions of quantum theory: the problem
of the quantum-to-classical transition. This has already been 
noted at the Solvay congress by Born (Jammer~1974):
\begin{quote}
\ldots how can it be understood that the trace of each $\alpha$-particle
[in the Wilson chamber] appears as an (almost) straight line \ldots?
\end{quote}
The problem becomes especially transparent in the correspondence
between Born and Einstein. As Einstein wrote to Born:
\begin{quote}
Your opinion is quite untenable. It is in conflict with the
principles of quantum theory to require that the $\psi$-function
of a ``macro"-system be ``narrow" with respect to the macro-coordinates
and momenta. Such a demand is at variance with the superposition 
principle for $\psi$-functions.
\end{quote}
More details can be found in Giulini {\em et al.}~(1996). 

During the last 25 years it became clear that a crucial role
in this quantum-to-classical transition is played by the natural
environment of a quantum system. Classical properties emerge
in an irreversible manner through the unavoidable interaction
with the ubiquitous degrees of freedom of the environment -- 
a process known as {\em decoherence}. This is the topic of our
contribution. Decoherence can be quantitatively understood in many
examples, and it has been observed in experiments. 
A comprehensive review with an (almost) exhaustive list of
references is Giulini {\em et al.} (1996) to which we refer for more details.
Reviews have also been given by Zurek (1991) and Zeh (1997),
see in addition the contributions by d'Espagnat, Haroche, and
Omn\`{e}s to this volume.  

Section~2 contains a general introduction to the essential mechanisms
of decoherence. The main part of our contribution are the examples
presented in Section~3. First, from special cases a detailed understanding
of how decoherence acts can be gained. Second, we choose examples
from all branches of physics to emphasise the encompassing aspect
of decoherence. Finally, Section~4 is devoted to interpretation:
Which conceptual problems are solved by decoherence, and which
issues remain untouched? We also want to relate some aspects to
other contributions at this conference and to perform an outlook
onto the ``Quantum Future" of decoherence.
 

\section{Decoherence: General Concepts}

\noindent
Let us now look in some detail at the general mechanisms and phenomena
which
arise from the interaction of a (possibly macroscopic) quantum system
with its environment. Needless to say that all effects depend on
the strength of the coupling between the considered degree of freedom
and the rest of the world. It may come as a surprise, however, that even
the
scattering of a single photon or the gravitational interaction
with far-away objects can lead to dramatic effects. To some
extent analogous outcomes can already be found in classical theory
(remember Borel's
example of the influence of a small mass, located on Sirius, on the
trajectories
of air molecules here on earth), but in quantum theory we
encounter as a new characteristic phenomenon the destruction of
coherence.
In a way this constitutes a violation of the superposition principle:
certain states can no longer be observed, although these
would be allowed by the theory. Ironically, this ``violation" is a
consequence of 
the assumed unrestricted validity of the superposition prinicple.
The destruction of coherence -- and
 to some extent the creation of classical properties -- was already
realized by the pioneers of quantum mechanics (see, for example,
Landau 1927, Mott 1929, and
Heisenberg 1958). In these early days (and even later), the influence
of the environment was mainly viewed as a kind of disturbance, 
exerted by a (classical) force. Even today such pictures are widespread,
although they are quite obviously incompatible with quantum theory.

The fundamental importance of decoherence in the macroscopic domain
seems to
have gone unnoticed for nearly half a century. Beginning with the work
of Zeh (1970), decoherence phenomena came under closer scrutiny in
the following two decades, first theoretically
 (K\"ubler and Zeh~1973, Zurek~1981, Joos and Zeh~1985,
Kiefer~1992, Omn\`{e}s~1997, and others),
now also experimentally (Brune {\em et al.}~1996).

\subsection{Decoherence and Measurements}

The mechanisms which are most important for the study of decoherence 
phenomena have much in common with those arising in the quantum
theory of measurement. We shall discuss below the interaction of a mass 
point with its environment in some detail. If the mass point is 
macroscopic -- a grain
of dust, say -- scattering of photons or gas molecules will transfer
information about the position of the dust grain into the
environment. In this sense, the position of the grain is ``measured" in
the
course of this interaction: The state of the rest of the universe (the
photon, at least) attains information about its position.

Obviously, the back-reaction (recoil) will be negligible in such a case,
hence we have a so-called ``ideal" measurement: Only the state of the
``apparatus" (in our case the photon) will change appreciably.
Hence there is no disturbance whatsoever of the measured system,
in striking conflict to early interpretations of quantum theory.

The quantum theory for ideal measurements was already formulated by
von Neumann in 1932 and is well-known, so here we need only to recall
the
essentials. Let the states of the measured system which are
discriminated by the apparatus be denoted by $|n\rangle$, then an
appropriate interaction Hamiltonian has the form
\be H_{int} =\sum_n|n\rangle\langle n| \otimes\hat{A}_n\ . \ee
The operators $\hat{A}_n$, acting on the states of the apparatus, are
rather
arbitrary, but must of course depend on the ``quantum number" $n$.
Note that the measured ``observable" is thereby dynamically defined by
the system-apparatus interaction and there is no reason to introduce
it
axiomatically (or as an additional concept).
If the
measured system is initially in the state $|n\rangle$ and the device in
some initial state $|\Phi_0\rangle$,
the evolution according to the Schr\"odinger equation
with Hamiltonian (1) reads
\bea |n\rangle|\Phi_0\rangle \stackrel{t}{\longrightarrow}
     \exp\left(-\I H_{int}t\right)|n\rangle|\Phi_0\rangle
     &=& |n\rangle\exp\left(-\I\hat{A}_nt\right)|\Phi_0\rangle\nonumber
\\
     &=:& |n\rangle|\Phi_n(t)\rangle\ . \eea
The resulting apparatus states $|\Phi_n(t)\rangle$ are usually called
``pointer positions", although in the general case of decoherence,
which we want to study here, they do not need to correspond to any
states of actually present measurement devices. They are simply the
states
of the ``rest of the world". An analogon to (2) can also be
written down in classical physics. The essential new quantum features
now come into play when we consider a {\em superposition} of different
eigenstates (of the measured ``observable") as initial state. The
linearity of time evolution immediately leads to
\be \left(\sum_n c_n|n\rangle\right)|\Phi_0\rangle
    \stackrel{t}\longrightarrow\sum_n c_n|n\rangle
    |\Phi_n(t)\rangle\ . \ee
If we ask, what can be seen when observing the measured system {\em
after}
this process, we need -- according to the quantum rules -- to calculate
the density matrix $\rho_S$
 of the considered system which evolves according to
\be \rho_S=\sum_{n,m}c_m^*c_n|m\rangle\langle n|
    \stackrel{t}{\longrightarrow}\sum_{n,m}c_m^*c_n
    \langle\Phi_m|\Phi_n\rangle|m\rangle\langle n|\ .\ee
If the environmental (pointer) states are approximately orthogonal,
\be \langle\Phi_m|\Phi_n\rangle \approx \delta_{mn}\ , \ee
that is, in the language of measurement theory, the measurement process
allows to discriminate the states $|n\rangle$ from each other, the
density
matrix becomes approximately diagonal in this basis,
\be \rho_S \approx \sum_n|c_n|^2|n\rangle\langle n|
       \ . \ee
Thus, the result of this interaction is a density matrix which seems
to describe an ensemble of different outcomes $n$ with the
respective probabilities. One must be careful in analyzing its
interpretation,
however. This density matrix only corresponds to an {\it apparent}
ensemble,
not a genuine ensemble of quantum states.
What can safely be stated is the fact, that interference terms 
(non-diagonal elements) are gone, hence
the coherence present in the initial system state in (3) can no longer
 be observed. Is coherence really ``destroyed"? Certainly not. The
right-hand side
 of (3) still displays a superposition of different $n$. The
coherence
is only {\it delocalised} into the larger system. As is well known, any
interpretation of a superposition as an ensemble
of components can be disproved experimentally
by creating interference effects. The same is true for 
the situation described in (3). For
example, the evolution could {\it in principle} be reversed. Needless
to say that such a reversal is experimentally extremely difficult, but
the interpretation and consistency of a physical theory must not depend
on our present technical abilities. Nevertheless, one often finds
explicit
or implicit statements to the effect that the above processes are 
equivalent to the collapse of the wave function (or even solve
the measurement problem). Such statements are certainly unfounded. What
can safely be said, is that coherence between the subspaces of the
Hilbert space spanned by $|n\rangle$ can no longer be observed
at the considered system, {\it if} the
process described by (3) is practically irreversible.

The essential implications are twofold: First, processes of the kind
(3) do happen
frequently and unavoidably for all macroscopic objects. Second, these
processes are irreversible in practically all
realistic situtations. In a normal measurement process, the
interaction and the state of the apparatus are controllable to some
extent (for example, the initial state of the apparatus is known to the
experimenter). In the case of decoherence, typically the initial state
is not known in detail (a standard example is interaction with
thermal radiation), but the consequences for the local density
matrix are the same: If the environment is described by an ensemble,
each member of this ensemble can act in the way described above.

A complete treatment of realistic cases has to include the Hamiltonian
governing the evolution of the system itself (as well as that of the
environment). The exact dynamics of a subsystem is hardly manageable
 (formally it is given by a complicated integro-differential
equation, see Chapter 7 of Giulini {\em et al.} 1996).
Nevertheless, we can find important approximate solutions
in some simplifying cases, as we shall show below.

\subsection{Scattering Processes}

An important example of the above-mentioned approximations is given
by scattering processes. Here we can separate the internal motion of the
system and the interaction with the environment, if the duration of
a scattering process is small compared to the timescale of the internal 
dynamics. The equation of motion is then a combination of the usual
von Neumann equation (as an equivalent to the unitary Schr\"odinger
equation) and a contribution from scattering, which may be calculated by
means of an appropriate S-matrix,
\be \I\frac{\partial\rho}{\partial t}
   =\left. [H_{internal},\rho]+\I\frac{\partial\rho}{\partial t}
    \right\vert_{scatt.} \ . \ee
In many cases, a sequence of scattering processes,
which may individually be quite inefficient but occur in a large
number, leads to an exponential damping of non-diagonal elements,
such as
\be \left.\frac{\partial\rho_{nm}}{\partial t}\right\vert_{scatt.}
     =-\lambda\rho_{nm}(t) \ee
with
\be \lambda=\Gamma(1-\langle\Phi_0|S_m^{\dagger}S_n|\Phi_0\rangle) \
.\ee
Here, $\Gamma$ is the collision rate, and the scattering processes 
off the states $|n\rangle$ and $|m\rangle$
are described by their corresponding S-matrix.

\subsection{Superselection Rules}

Absence of interference between certain states, that is, non-observation
of
certain superpositions is often called a superselection rule. This term
was
coined by Wick, Wightman and Wigner in 1952 as a generalization of
the term ``selection rule".

In the framework of decoherence, we can easily see that superselection
rules
are induced by interaction with the environment. If interference
terms are destroyed fast enough, the system will always appear as a
mixture of
states from different {\it superselection sectors}. In contrast to
axiomatically postulated superselection rules (often derived from
symmetry arguments), superselection rules are never exactly valid in
this framework, but
only as an approximation, depending on the concrete situation. We shall
give some examples in the next Section.

\section{Decoherence: Examples}

In the following we shall illustrate some features of decoherence by 
looking at special cases from various fields of physics.
We shall start from examples in nonrelativistic
quantum mechanics and then turn to examples in quantum electrodynamics
and quantum gravity.

\subsection{Localisation of Objects}

Why do macroscopic objects always appear localised in space? Coherence 
between macroscopically different positions is destroyed {\it very}
rapidly
 because
of the strong influence of scattering processes. The formal description
may proceed as follows. Let $|x\rangle$ be the position eigenstate
of a macroscopic object, and $|\chi\rangle$ the state of the
incoming particle.
 Following the von Neumann scheme, the scattering of
such particles off an object located at position $x$ may be written as
\be |x\rangle|\chi\rangle \stackrel{t}{\longrightarrow}
    |x\rangle|\chi_x\rangle=|x\rangle S_x|\chi\rangle\ , \ee
where the scattered state may conveniently be calculated by means of
an appropriate S-matrix. For the more general initial state of a wave
packet we have then
\be \int\D^3x\ \varphi(x)|x\rangle|\chi\rangle
    \stackrel{t}{\longrightarrow}\int\D^3x\ \varphi(x)|x\rangle
     S_x|\chi\rangle\ , \ee
and the reduced density matrix describing our object changes
into
\be \rho(x,x')=\varphi(x)\varphi^*(x')
  \left\langle\chi|S_{x'}^{\dagger}S_x|\chi\right\rangle\ . \ee
Of course, a single scattering process will usually not resolve a small
distance, so
in most cases the matrix element on the right-hand side
 of (12) will be close to unity.
But if we add the contributions of many scattering processes, an
exponential damping of spatial coherence results:
\be \rho(x,x',t)= \rho(x,x',0)\exp\left\{-\Lambda t(x-x')^2\right\}\ .
\ee
The strength of this effect is described by a single parameter $\Lambda$
which may be called ``localisation rate" and is given by
\be \Lambda= \frac{k^2Nv\sigma_{eff}}{V}\ .\ee
Here, $k$ is the wave number of the incoming particles, $Nv/V$
the flux, and $\sigma_{eff}$ is of the order of the total cross section
(for details see Joos and Zeh 1985 or
Sect. 3.2.1 and Appendix 1 in Giulini {\em et al.} 1996). 
Some values of $\Lambda$ are given in the Table.

\begin{table}[htb]
\caption[ ]{Localisation rate $\Lambda$ in $\mbox{cm}^{-2}
\mbox{s}^{-1}$ for three sizes of ``dust particles" and various
types of scattering processes (from Joos and Zeh~1985).
This quantity measures how fast interference between different
positions disappears as a function of distance in the course of
time, see (13).}
\begin{flushleft}
\renewcommand{\arraystretch}{1.2}  
\begin{tabular}{l|lll}
\hline
{} & \ $a=10^{-3}\mbox{cm}$ &\  $a=10^{-5}\mbox{cm}$ &\
$a=10^{-6}\mbox{cm}$\\
{} & \ dust particle        &\  dust particle        &\ large molecule
\\ \hline
Cosmic background radiation &\ $10^{6}$& $10^{-6}$ &$10^{-12}$ \\
300 K photons &\ $10^{19}$ & $10^{12}$ & $10^6$ \\
Sunlight (on earth) &\ $10^{21}$  & $10^{17}$ & $10^{13}$ \\
Air molecules & \  $10^{36}$ & $10^{32}$ & $10^{30}$ \\
Laboratory vacuum & \ $10^{23}$ & $10^{19}$ & $10^{17}$\\
($10^3$ particles/$\mbox{cm}^3$) & {}&{}&\\
\hline
\end{tabular}
\renewcommand{\arraystretch}{1}
\end{flushleft}\end{table}

Most of the numbers in the table are quite large, showing the extremely
strong coupling of macroscopic objects, such as dust particles, to their
natural environment. Even in intergalactic space, the 3K background
radiation
cannot simply be neglected.

Let us illustrate the effect of decoherence for the case of a
superposition
of two wave packets. If their distance is ``macroscopic", then such
states
are now usually called ``Schr\"odinger cat states". Fig. 1a shows the
corresponding density matrix, displaying four peaks, two along the main
diagonal and two off-diagonal contributions representing coherence
between the
two parts of the extended wave packet.

\begin{figure}
\epsfig{file=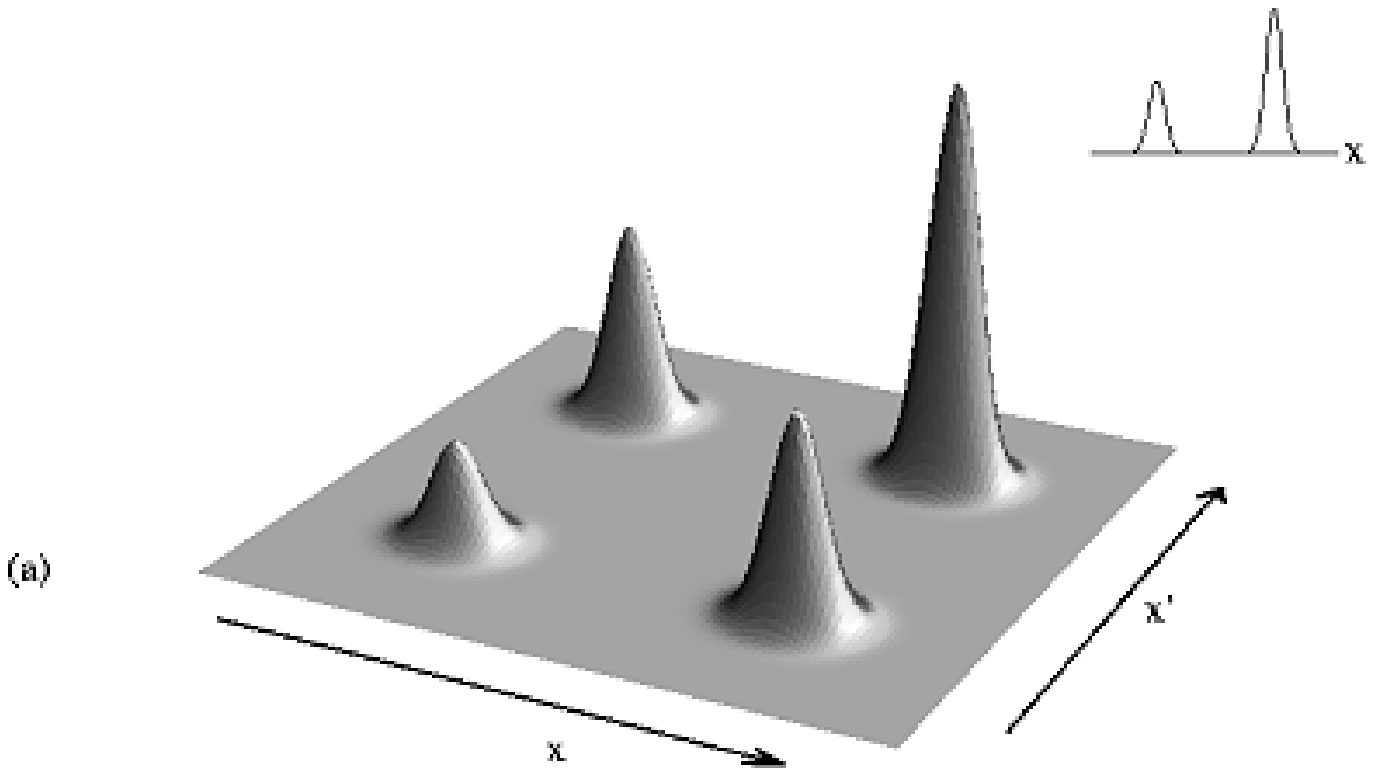}
\epsfig{file=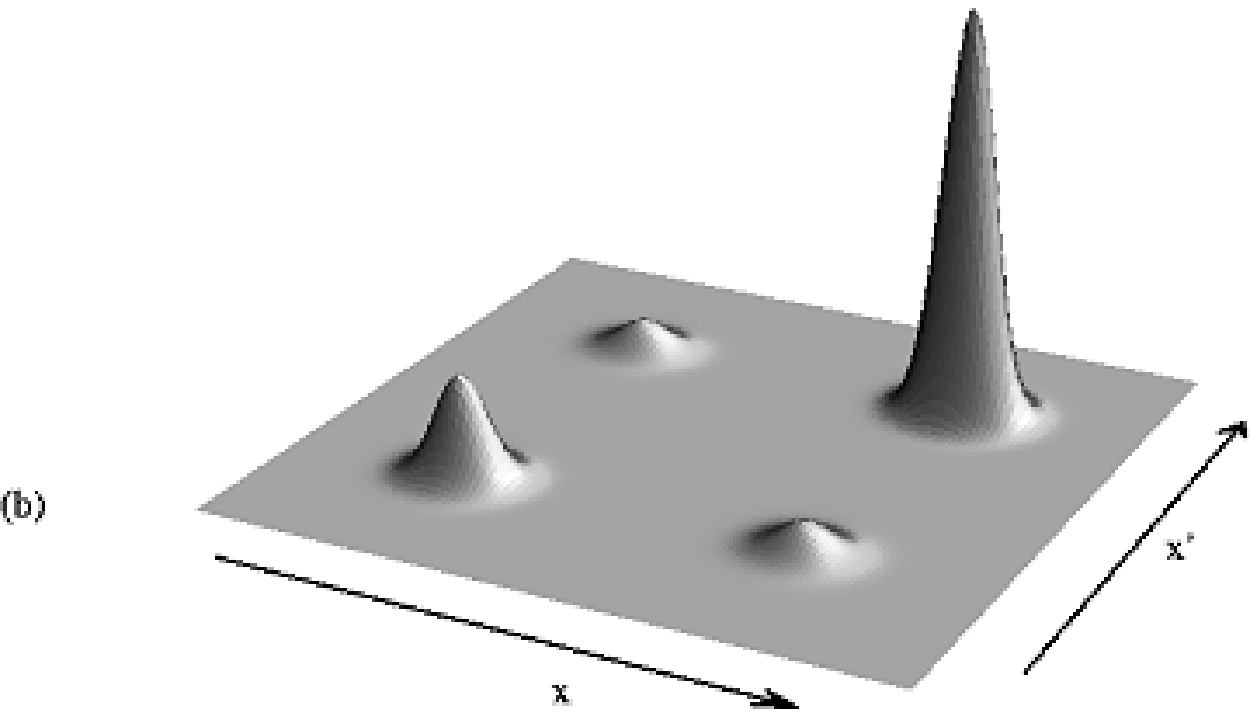}
\caption[]{{\bf (a)} Density matrix of a superposition of two Gaussian
wave packets. The wave function is shown in the inset. Coherence
between the two parts of the wave function is represented by the
two off-diagonal peaks. {\bf (b)} The density matrix after interference
is partially destroyed by decoherence. The position distribution,
along the diagonal, is not changed appreciably.}
\end{figure}

Decoherence according to (13) leads to damping of off-diagonal terms,
whereas
the peaks near the diagonal are not affected appreciably
(this is a property of an ideal measurement). Thus the density matrix
develops into a mixture of two packets, as shown in Fig. 1b. 

The same effect can be described by using the Wigner function, which
is given in terms of the density matrix as
\be W(x,p)=\frac{1}{\pi}\int_{-\infty}^{\infty}
    \D y\ \E^{2\I py}\rho(x-y,x+y)\ .\ee

\begin{figure}
\epsfig{file=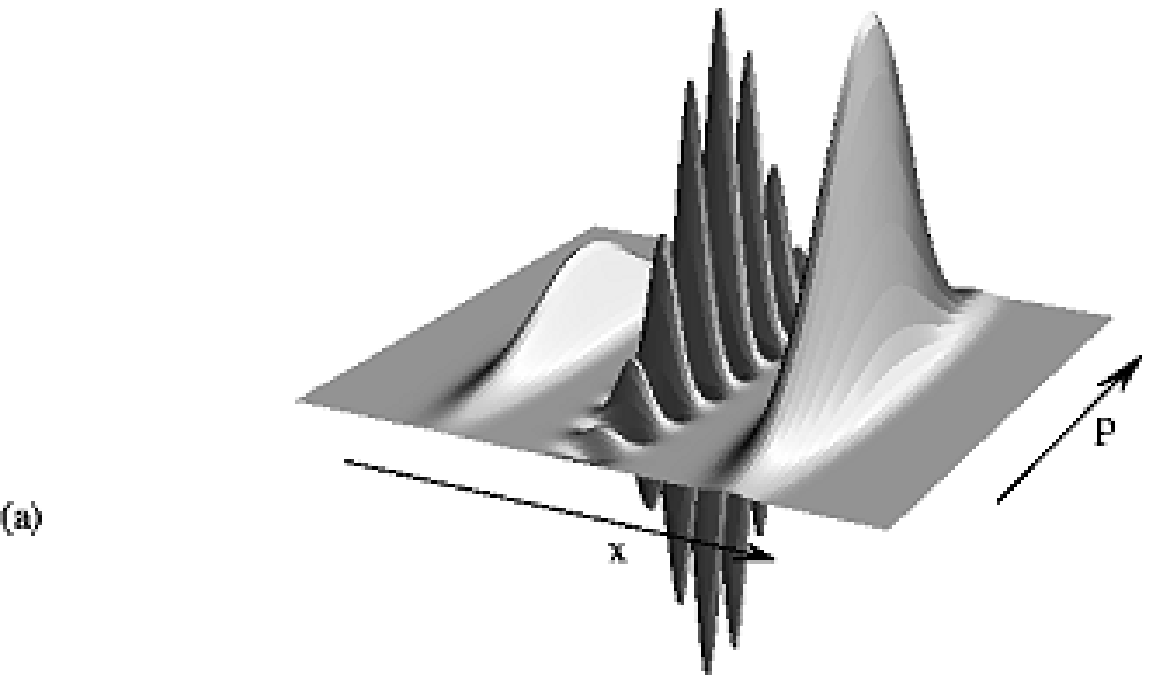}
\epsfig{file=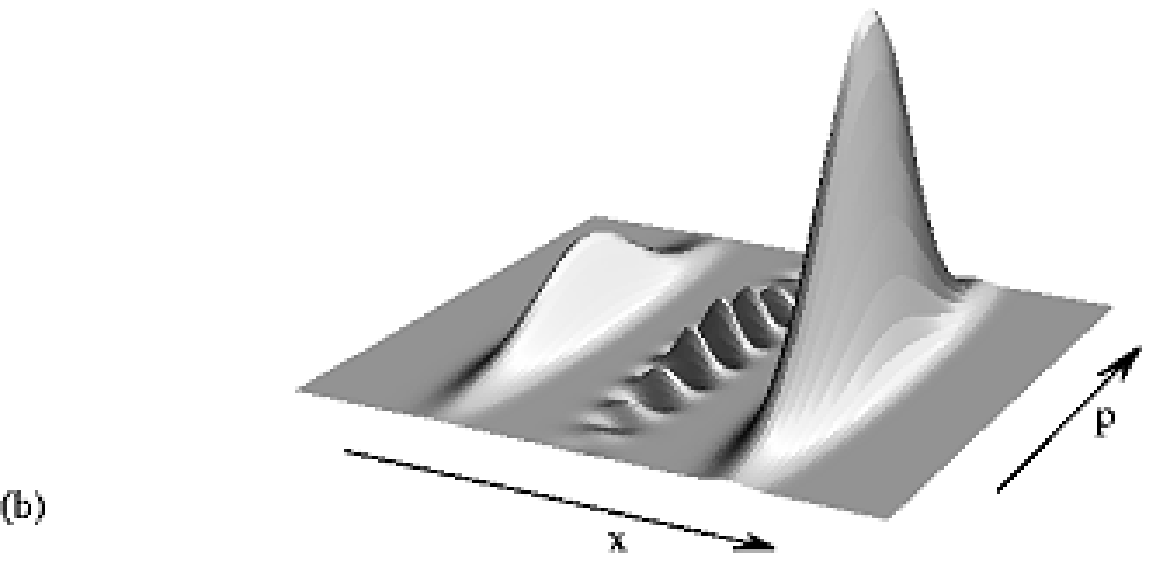}

\caption[]{The Wigner function equivalent to the density matrices
shown in Fig. 1. {\bf (a)} represents the superposition of
two Gaussian wave packets. Strong oscillations together with negative
values indicate coherence between the two wave packets. {\bf (b)}
 oscillations are partially damped by decoherence.}
\end{figure}

A typical feature of the Wigner function are the oscillations occurring
for
``nonclassical" states, as can be seen in Fig. 2a.  These oscillations
are
damped by decoherence, so that the Wigner function looks more and more
like
a classical phase space distribution (Fig. 2b). One should keep in mind,
however, that the Wigner function is only a useful calculational tool
and does not describe
a genuine phase space distribution of particles (which do not exist in
quantum theory). 

The following figures show the analogous situation for an
eigenstate of a harmonic oscillator.

\begin{figure}
\epsfig{file=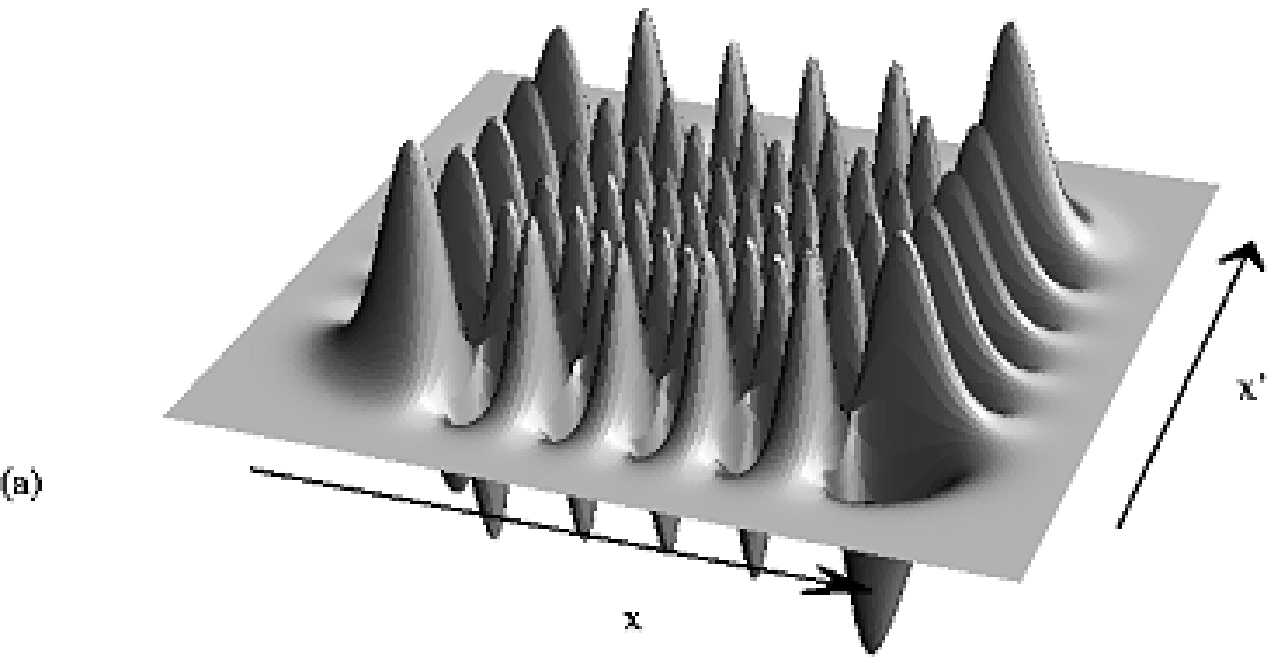}
\epsfig{file=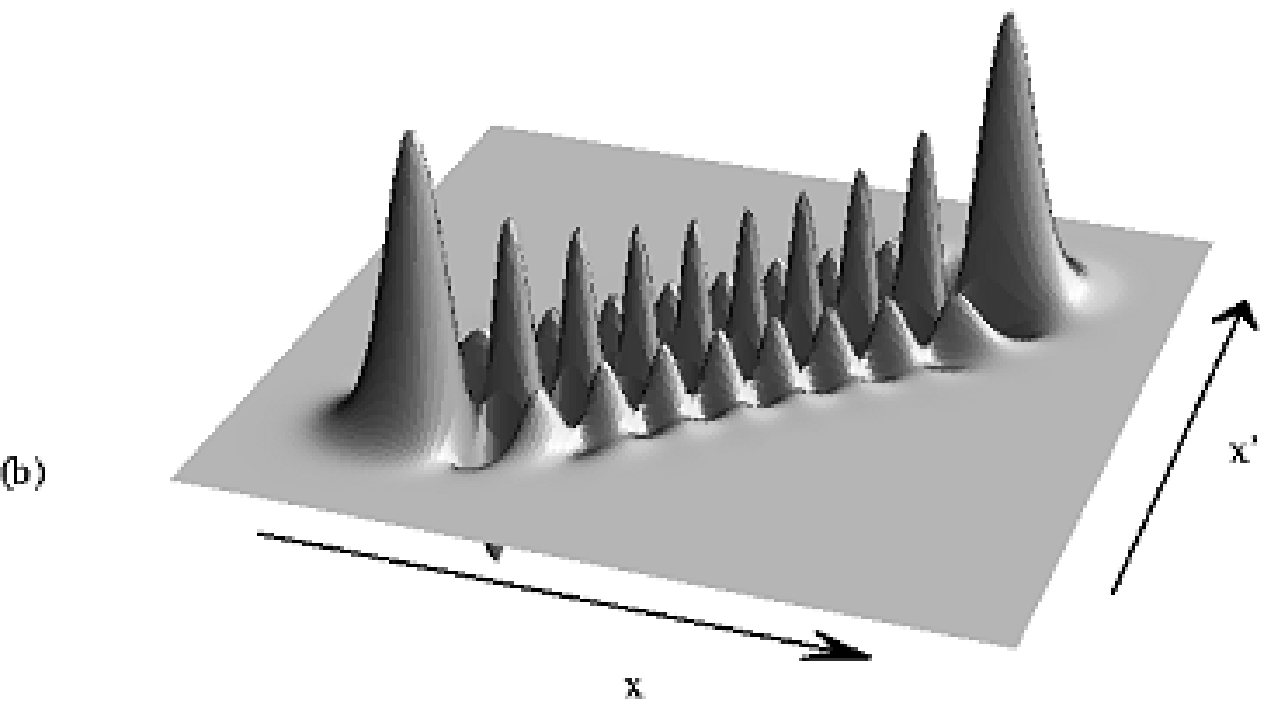}
\caption[]{{\bf (a)}Density matrix of an energy eigenstate of a
harmonic oscillator for {\it n}=9 in the position representation.
{\bf (b)} Non-diagonal terms are damped by decoherence.}
\end{figure}

\begin{figure}
\epsfig{file=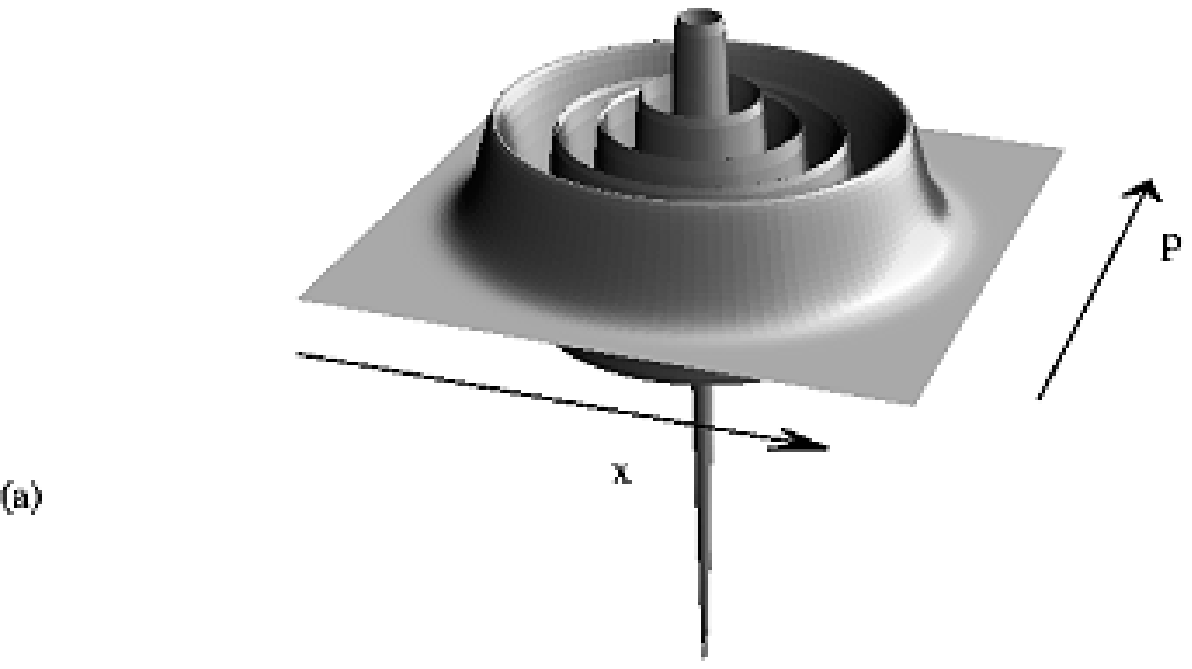}
\epsfig{file=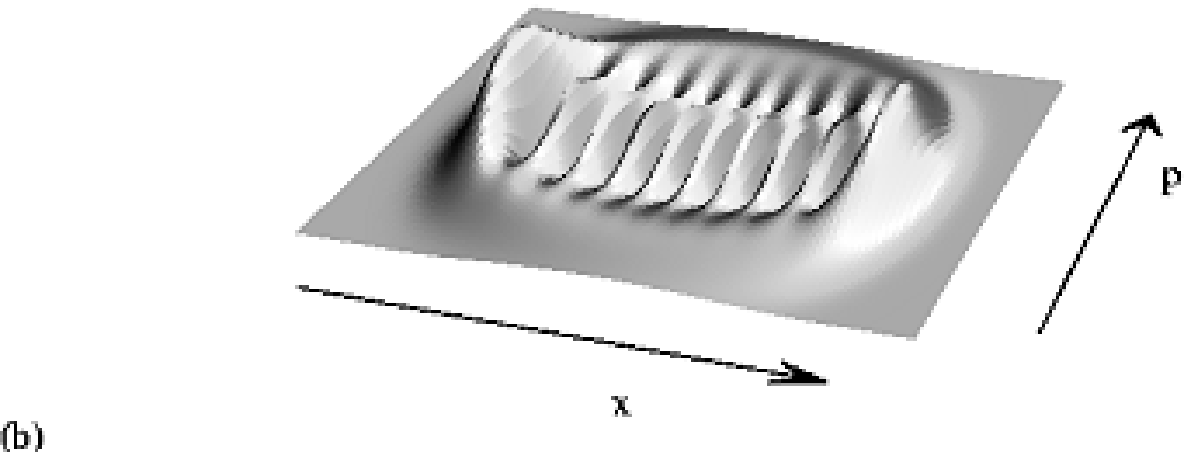}
\caption[]{{\bf (a)}Wigner function for an energy eigenstate of a
harmonic oscillator with {\it n}=9. The figure shows strong
oscillations indicating the non-classical character of this pure state. 
{\bf (b)} Decoherence acts like diffusion in this representation. Note
that the symmetry between position and momentum is broken by the
interaction with the environment.}
\end{figure}

Combining the decohering effect of scattering processes with the
internal dynamics
of a ``free" particle leads (as in (7)) 
 to a Boltzmann-type master equation which in one dimension is of the form
\be \I\frac{\partial\rho}{\partial t} =
   [H,\rho]-\I\Lambda[x,[x,\rho]]  \ee
and reads explicitly 
\be \I\frac{\partial\rho(x,x',t)}{\partial t}=
  \frac{1}{2m}\left(\frac{\partial^2}{\partial x'^2}
  -\frac{\partial^2}{\partial x^2}\right)\rho
  -\I\Lambda(x-x')^2\rho\ . \ee
Solutions can easily be found for these equations
(see Appendix~2 in Giulini {\em et al.}~1996).
Let us look at one typical quantum property, the coherence length.
According to the Schr\"odinger equation, a free wave packet would
spread,
thereby increasing its size and extending its coherence properties over
a larger region of space. Decoherence is expected to counteract this
behaviour and reduce the coherence length. This can be seen in the
solution shown in Fig. 5, where the time dependence of the coherence
length (the width of the density matrix in the off-diagonal direction)
is
plotted for a truly free particle
 (obeying a Schr\"odinger equation) and also for increasing strength of
decoherence. For large times the spreading of the
wave packet no longer occurs and the coherence length always decreases
proportional to $1/\sqrt{\Lambda t}$.

\begin{figure}
\epsfig{file=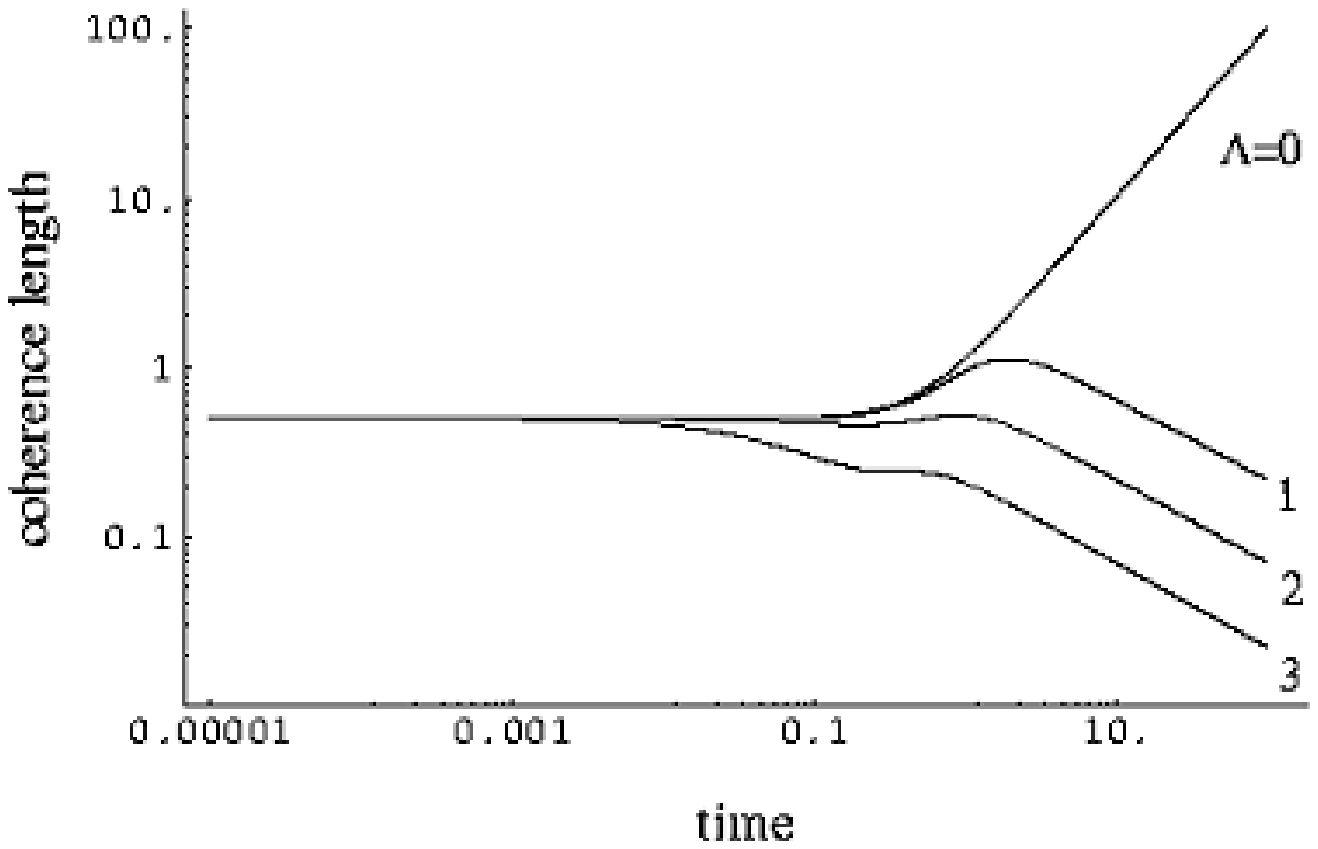,scale=0.5}
\caption[]{Time dependence of coherence length. It is a measure of
the spatial extension over which the object can show interference
effects. Except for zero coupling ($\Lambda=0$), the coherence
length always decreases for large times.}
\end{figure}

Is all this just the effect of thermalization? There are several models
for
the quantum analogue of Brownian motion, some of which are even older
than
the first decoherence studies. Early treatments did not, however, 
draw a distinction
between decoherence and friction. As an example, consider the equation
of motion derived by Caldeira and Leggett (1983),
\be \I\frac{\partial\rho}{\partial t}= [H,\rho]
   +\frac{\gamma}{2}[x,\{p,\rho\}]-\I m\gamma k_BT[x,[x,\rho]] \ee
which reads in one space dimension of a ``free" particle
\bea \I\frac{\partial\rho(x,x',t)}{\partial t}&=& \left[
  \frac{1}{2m}\left(\frac{\partial^2}{\partial x'^2}
  -\frac{\partial^2}{\partial x^2}\right)-\I\Lambda(x-x')^2\right.
   \nonumber\\
   & & \left. +\I\gamma(x-x')\left(\frac{\partial}{\partial x'}
   -\frac{\partial}{\partial x}\right)\right]
  \rho(x,x',t)\ , \eea
where $\gamma$ is the damping constant and here $\Lambda=m\gamma k_BT$.
If one compares the effectiveness of the two terms representing
decoherence and relaxation, one finds that their ratio is given by
\be \frac{\mbox{decoherence rate}}{\mbox{relaxation rate}}
   =mk_BT(\delta x)^2\propto \left(\frac{\delta x}{\lambda_{th}}
     \right)^2 \ , \ee
where $\lambda_{th}$ denotes the thermal de~Broglie wavelength.
This ratio has for a typical macroscopic situation 
($m=1 \mbox{g}$, $T=300\mbox{K}$, $\delta x=1 \mbox{cm}$) the enormous
value of
about $10^{40}$! This shows that in these cases decoherence is {\em far
more important} than dissipation.

Not only the centre-of-mass position of dust particles becomes
``classical" via 
decoherence. The spatial structure of molecules represents another most
important example. Consider a simple model of a chiral molecule (Fig.
6).

\begin{figure}
\epsfig{file=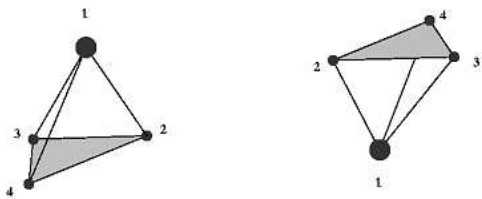,scale=0.5}
\caption[]{Typical structure of an optically active, chiral molecule.
Both versions are mirror-images of each other and are not connected
by a proper rotation, if the four elements are different.}
\end{figure}

Right- and left-handed versions both have a rather well-defined spatial
structure, whereas the ground state is - for symmetry reasons -
a superposition of both chiral states. These chiral configurations are
usually separated by a tunneling barrier (compare Fig. 7) which is so
high that under
normal circumstances tunneling is very improbable, as was already
shown by Hund in 1929. But this alone does not explain why chiral
molecules are never found in energy eigenstates! 

\begin{figure}
\epsfig{file=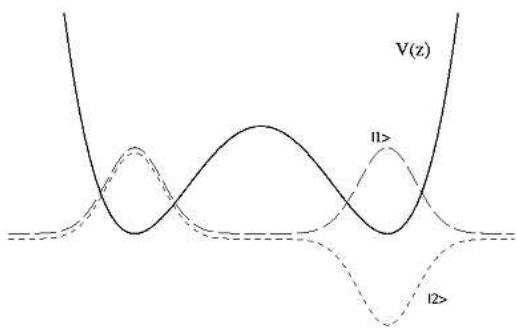,scale=0.5}
\caption[]{Effective potential for the inversion coordinate in a model
for a chiral molecule and the two lowest-lying eigenstates. The
ground state is symmetrically distributed over the two wells. Only
linear combinations of the two lowest-lying states are localised and
correspond to a classical configuration.}
\end{figure}

In a simplified model with low-lying nearly-degenerate eigenstates
$|1\rangle$
and $|2\rangle$, the right- and left-handed configurations may be given
by
\bea |L\rangle &=& \frac{1}{\sqrt{2}}(|1\rangle+|2\rangle)\nonumber\\
     |R\rangle &=& \frac{1}{\sqrt{2}}(|1\rangle-|2\rangle)\ . \eea
Because the environment recognises the spatial structure, only chiral
states are stable against decoherence,

\be |R,L\rangle|\Phi_0\rangle \stackrel{t}{\longrightarrow}
    |R,L\rangle|\Phi_{R,L}\rangle\ . \ee

Additionally, transitions between spatially oriented states are
suppressed by the quantum Zeno effect, described below.

\subsection{Quantum Zeno Effect}
The most dramatic consequence of a strong measurement-like interaction
of
a system with its environment is the quantum Zeno effect. It has been
discovered
several times and is also sometimes called ``watchdog effect" or
``watched pot
behaviour",
although most people now use the term Zeno effect.
It is surprising only if one sticks to a classical picture
where observing a system and just verifying its state should have no
influence on it. Such a prejudice is certainly formed by our
everyday experience, where observing things in our surroundings does
not change their behaviour. As is known since the early times of quantum
theory, observation can drastically change the observed system.

The essence of the Zeno effect can easily be shown as follows. Consider
the
``decay" of a system which is initially prepared in the ``undecayed"
state $|u\rangle$. The probability to find the system undecayed, i.e.,
in the
same state $|u\rangle$ at time $t$ is for small time intervals given by
\bea P(t) &=& |\langle u|\exp(-\I Ht)|u\rangle|^2 \nonumber\\
          &=& 1-(\Delta H)^2t^2+ {\cal O}(t^4) \eea
with
\be (\Delta H)^2= \langle u|H^2|u\rangle - \langle u|H|u\rangle^2 \ .\ee
If we consider the case of $N$ measurements in the interval $ [0,t]$,
the
non-decay probability is given by
\be P_N(t)\approx\left[1-(\Delta H)^2\left(\frac{t}{N}\right)^2
    \right]^N > 1-(\Delta H)^2t^2= P(t)\ . \ee
This is always larger than the single-measurement probability
given by (23).
In the limit of arbitrary dense measurements, the system no longer
decays,
\be P_N(t)= 1-(\Delta H)^2\frac{t^2}{N}+\ldots
    \stackrel{N\to\infty}{\longrightarrow}1\ .\ee
Hence we find that repeated measurements can completely hinder the
natural
evolution of a quantum system. Such a result is clearly quite distinct
from what is observed for classical systems. Indeed, the paradigmatic
example
for a classical stochastic process, exponential decay,
\be P(t)=\exp(-\Gamma t)\ , \ee
is not influenced
by repeated observations, since for $N$ measurements we simply have
\be P_N(t)=\left(\exp\left(-\Gamma\frac{t}{N}\right)\right)^N
          =\exp(-\Gamma t)\ . \ee
So far we have treated the measurement process in our discussion of the
Zeno
effect in the usual way by assuming a collapse of the system state onto
the subspace corresponding to the measurement result. Such a treatment
can be
extended by employing a von Neumann model for the measurement process,
e.g., by coupling a pointer to a two-state system. A simple toy model is
given
by the Hamiltonian
\bea H&=& H_0+H_{int}\nonumber\\
   &=& V(|1\rangle\langle 2|+|2\rangle\langle 1|)
   +E|2\rangle\langle 2|+\gamma\hat{p}(|1\rangle\langle 1|
   -|2\rangle\langle 2|)\ , \eea
where transitions between states $|1\rangle$ and $|2\rangle$ (induced by
the ``perturbation" {\it V}) are monitored by a pointer (coupling
constant $\gamma$). This model already shows all the typical features
mentioned above (see Fig.~8).

\begin{figure}
\epsfig{file=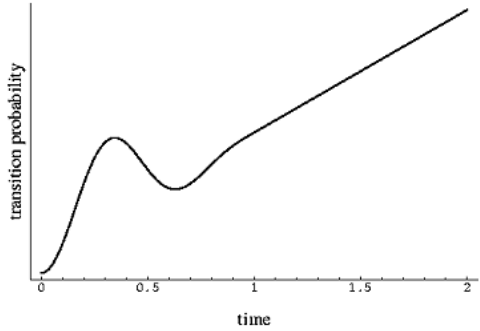,scale=0.5}
\caption[]{Time dependence of the probability of finding state
$|2\rangle$ if the system was prepared in $|1\rangle$ at $t=0$
under continuous coupling to a meter.}
\end{figure}

The transition probability starts for small times always quadratically,
according to the general result (23). For times, where the pointer 
resolves the two states, a behaviour similar to that found for Markow
processes appears: The quadratic time-dependence changes to a linear
one.

\begin{figure}
\epsfig{file=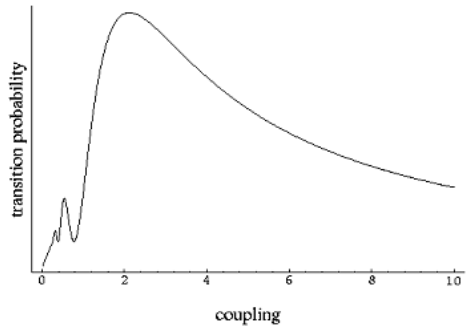,scale=0.5}
\caption[]{Probability of finding state $|2\rangle$ at a fixed time
as a function of the coupling to the meter.}
\end{figure}

Fig. 9 displays the transition probability as a function of the coupling
strength. For strong couling the transitions are suppressed. This
clearly shows the dynamical origin of the Zeno effect.

An extension of the above model allows an analysis of the transition
from
the Zeno effect to master behaviour (described by transition rates as was
first studied by Pauli in 1928). It can be shown that for many (micro-)
states which are not sufficiently resolved by the environment, Fermi's
Golden
Rule can be recovered, with transition rates which are no longer
reduced by the Zeno effect. Nevertheless, interference between
macrostates
is suppressed very rapidly (Joos 1984).


\subsection{An example from Quantum Electrodynamics}

The occurrence of decoherence is a general phenomenon in quantum
theory and is by no means restricted to nonrelativistic quantum
mechanics. The following two sections are devoted to
decoherence in QED and quantum gravity. It is obvious that the
discussion there is technically more involved, and
these areas are therefore less
studied. However, interesting physical aspects turn out from an 
understanding of decoherence in this context.

Two situations are important for decoherence in QED, which are, however,
two sides of the same coin (Giulini {\em et al.}~1996):
\begin{itemize}
\item ``Measurement" of charges by fields;
\item ``Measurement" of fields by charges.
\end{itemize}
In both cases the focus is, of course, on quantum entanglement
between states of charged fields with the electromagnetic field,
but depending on the given situation, the roles of ``relevant"
and ``irrelevant" parts can interchange.

Considering charges as the relevant part of the total system,
it is important to note that every charge is naturally correlated
with its Coulomb field. This is a consequence of the Gau\ss\ constraint.
 Superpositions of charges are therefore
nonlocal quantum states. This lies at the heart of the charge
superselection rule: {\em Locally} no such superpositions can be
observed, since they are decohered by their entanglement
with the Coulomb fields. 

A different, but related, question is how far, for example, a wave
packet
with one electron can be spatially separated and coherently
combined again. Experiments show that this is possible over
distances in the millimetre range. 
Coulomb fields act {\em reversibly}
and cannot prevent the parts of the electronic
wave packet from coherently recombining. Genuine, irreversible,
decoherence is achieved by emission of real photons. 
A full QED calculation for this decoherence process is elusive,
although estimates exist (Giulini {\em et al.}~1996).

Let us now focus on the second part, the decoherence of the
electromagnetic field through entanglement with charges.
A field theoretic calculation was first done by Kiefer (1992)
in the framework of the functional Schr\"odinger picture for
scalar QED. One particular example discussed there is the superposition
of a semiclassical state for the electric field pointing upwards
with the analogous one pointing downwards,
\be |\Psi\rangle \approx |E\rangle|\chi_E\rangle
      +|-E\rangle|\chi_{-E}\rangle\ , \ee
where the states $|\chi\rangle$ depend on $E$
and represent states for the charged field. The state $\Psi$ is an
approximate solution to the full functional Schr\"odinger equation.
The corresponding reduced density matrix for the electric field
shows four ``peaks", in analogy to the case shown in Fig.~1.
We take as an example a Gaussian state for the charged fields,
representing an adiabatic vacuum state.
Integrating out the charged fields from (30), the interference
terms (non-diagonal elements) become suppressed, while the 
probabilities (diagonal elements) are only a slightly changed,
corresponding to an almost ideal measurement.
In particular, one gets for the non-diagonal elements
$\rho_{\pm}=\rho_{\mp}^*$:
\be \rho_{\pm}\stackrel{t}{\longrightarrow}\rho_{\pm}(D_V+D_{PC})\ ,\ee
where $D_V$ is the contribution from vacuum polarisation
(having a reversible effect like the Coulomb field) and
$D_{PC}$ is the contribution from particle creation
(giving the typical irreversible behaviour of decoherence).
The explicit results are
\bea D_V &=&  \exp\left(-\frac{Vt}{256\pi^2}
  \frac{(eE)^3}{m^2+(eEt)^2}- \frac{V(eE)^2}{256\pi^2m}
  \arctan\frac{eEt}{m}\right) \nonumber\\
         &\stackrel{t\gg m/eE}{\longrightarrow} &
                              \exp\left
  (-\frac{Ve^2E^2}{512\pi m}\right) \label{rhopm} \eea
and
\be D_{PC} =\exp\left(-\frac{Vte^2E^2}{4\pi^2}\E^{-\pi m^2/eE}
           \right)\ . \ee
Here, $m$ is the mass of the charged field, and $V$ is the 
volume; the system has to be enclosed in a finite box to
avoid infrared singularities. There is a critical field strength
$E_c\equiv m^2/e$, above which particle creation is important
(recall Schwinger's pair creation formula). For $E<E_c$, the usual
irreversible decoherence is negligible and only the contribution
from vacuum polarisation remains. On the other hand, for $E>E_c$,
particle creation is dominating, and one has
\be \left\vert\frac{\ln D_V}{\ln D_{PC}}\right\vert
    \propto \frac{\pi^2}{128 mt}\stackrel{t}{\longrightarrow} 0
   \ . \ee
Using the influence functional method, the same result
was found by Shaisultanov (1995a). He also studied the same situation
for fermionic QED and found a somewhat stronger effect for
decoherence, see Shaisultanov (1995b). The above result is 
consistent with the results of Habib {\em et al.} (1996) who also found
that decoherence due to particle creation is most effective.

\subsection{Decoherence in Gravity Theory}

In the traditional Copenhagen interpretation of quantum theory,
an a priori classical part of the world is assumed to exist from
the outset. Such a structure is there thought to be necessary for
the ``coming into being" of observed measurement results (according to
John Wheeler, only an observed phenomenon is a phenomenon).
The programme of decoherence, on the other hand, demonstrates
that the emergence of classical properties can be understood
within quantum theory, without a classical structure given a priori.
The following discussion will show that this also holds for the
structure that one might expect to be the most classical --
spacetime itself. 

In quantum theories of the gravitational field, no classical
spacetime exists at the most fundamental level. Since it is 
generally assumed that the gravitational field has to be quantised,
the question again arises how the corresponding classical
properties arise. 

Genuine quantum effects of gravity are expected to occur for scales
of the order of the Planck length $\sqrt{G\hbar/c^3}$. It is
therefore argued that the spacetime structure at larger scales
is automatically classical. However, this Planck scale argument
is as insufficient as the large mass argument in the evolution
of free wave packets. As long as the superposition principle is
valid (and even superstring theory leaves this untouched),
superpositions
of different metrics should occur at any scale. 

The central problem can already be demonstrated in a simple
Newtonian model. Following Joos (1986), we consider a cube of length
$L$ containing a homogeneous gravitational field with a
quantum state $\psi$ such that at some initial time $t=0$
\be  \vert\psi\rangle =c_1\vert g\rangle +c_2\vert g'\rangle,
    \; g\neq g'\ , \lb{su} \ee
where $g$ and $g'$ correspond to two different field strengths.
A particle with mass $m$
in a state $\vert\chi\rangle$, which moves through this volume,
``measures" the value of $g$, since its trajectory depends on the
metric:
\be \vert \psi\rangle\vert\chi^{(0)}\rangle \to
    c_1\vert g\rangle\vert\chi_g(t)\rangle
    +c_2|g'\rangle|\chi_{g'}(t)\rangle\ .
  \ee
This correlation destroys the coherence between $g$ and $g'$,
and the reduced density matrix can be estimated to assume the
following form after many such interactions are taken into
account:
\be \rho(g,g',t) =\rho(g,g',0)\exp\left(-\Gamma t(g-g')^2\right),
       \lb{rhog} \ee
 where
 \[ \Gamma =nL^4\left(\frac{\pi m}{2k_{{B}}T}\right)^{3/2} \]
 for a gas with particle density $n$ and temperature $T$.
 For example, air under ordinary conditions, and $L=1\mbox{ cm}$,
 $t=1\mbox{ s}$ yields a remaining coherence width of
 $\Delta g/g\approx 10^{-6}$. 

Thus, matter does not only tell space to curve but also to
behave classically. This is also true in full quantum gravity.
Although such a theory does not yet exist, one can discuss this
question within present approaches to quantum gravity.
In this respect, canonical quantum gravity fully serves this purpose
(Giulini {\em et al.} 1996).
Two major ingredients are necessary for the emergence of a classical
spacetime:
\begin{itemize}
\item A type of Born-Oppenheimer approximation for the gravitational
 field. This gives a semiclassical state for the 
gravitational part and a Schr\"odinger equation for the matter part
in the spacetime formally defined thereby.\footnote{More precisely,
also some gravitational degrees of freedom (``gravitons") 
must be adjoined to the matter degrees of freedom obeying the
Schr\"odinger equation.}
 Since also superstring
theory should lead to this level in some limit, the treatment within
canonical gravity is sufficient.
\item The quantum entanglement of the gravitational field with
irrelevant degrees of freedom (e.g. density perturbations)
leads to decoherence for the gravitational field. During this process,
states become distinguished that have a well-defined time
(which is absent in full quantum gravity). This symmetry breaking
stands in full analogy to the symmetry breaking for parity in the
case of chiral molecules, see Figs.~6 and~7.
\end{itemize}
The division between relevant and irrelevant degrees of freedom
can be given by the division between semiclassical degrees of freedom
(defining the ``background") and others. For example, the
relevant degrees of freedom may be the scale factor $a$ of
a Friedmann universe containing a global scalar field, $\phi$,
like in models for the inflationary universe. The irrelevant 
degrees of freedom may then be given by small perturbations
of these background variables, see Zeh (1986). 
Explicit calculations then yield a large degree of classicality
for $a$ and $\phi$ through decoherence (Kiefer 1987).
It is interesting to note that the classicality for $a$
is a necessary prerequisite for the classicality of $\phi$.

Given then the (approximate) classical nature of the spacetime
background, decoherence plays a crucial role for the emergence of
classical density fluctuations serving as seeds for galaxies and
clusters of galaxies (Kiefer, Polarski, and Starobinsky~1998):
In inflationary scenarios, all structure emerges from quantum
fluctuations of scalar field and metric perturbations.
If these modes leave the horizon during inflation, they become
highly squeezed and enter as such the horizon in the radiation
dominated era. Because of this extreme squeezing,
the field amplitude basis becomes a ``quantum nondemolition variable",
i.e. a variable that -- in the Heisenberg picture -- commutes
at different times. Moreover, since squeezed states are extremely
sensitive to perturbations through interactions with other fields,
the field amplitude basis becomes a perfect pointer basis
by decoherence. For these reasons, the fluctuations observed
in the microwave background radiation are classical stochastic
quantities; their quantum origin is exhibited only in the Gaussian
nature of the initial conditions.

Because the gravitational field universally interacts with all
other degrees of freedom, it is the first quantity (at least
its ``background part") to become classical. Arising from the
different types of interaction, this gives rise to the
following {\em hierarchy of classicality}:
\begin{center}
\fbox{Gravitational background variables}\\
$\downarrow$ \\
\fbox{Other background variables}\\
$\downarrow$\\ 
\fbox{Field modes leaving the horizon}\\
$\downarrow$\\
\fbox{Galaxies, clusters of galaxies}\\
$\downarrow$\\
\ldots
\end{center}
It must be emphasised that decoherence in quantum gravity is
not restricted to cosmology. For example, a superposition of
black and white hole may be decohered by interaction with
Hawking radiation (Demers and Kiefer~1996). However, this only
happens if the black holes are in semiclassical states.
For virtual black-and-white holes no decoherence, and therefore
no irreversible behaviour, occurs.

\section{Interpretation}

The discussion of the examples in the previous Section clearly
demonstrates the {\em ubiquitous} nature of decoherence -- it is
simply not consistent to treat most systems as being isolated.
This can only be assumed for microscopic systems such as atoms
or small molecules. 

 In principle, decoherence could have been studied already in the early
days of quantum mechanics and, in fact, the contributions of
Landau, Mott, and Heisenberg at the end of the twenties can be
interpreted
as a first step in this direction. Why did one not go further
at that time? One major reason was certainly the advent of the
``Copenhagen doctrine" that was sufficient to apply the formalism
of quantum theory on a pragmatic level. In addition, the imagination
of objects being isolable from their environment was so deeply
rooted since the time of Galileo, that the {\em quantitative} aspect
of decoherence was largely underestimated. This quantitative
aspect was only born out from detailed calculations, some of which
we reviewed above. Moreover, direct experimental verification
was only possible quite recently.

What are the achievements of the decoherence mechanism?
Decoherence can certainly explain why and how {\em within}
quantum theory certain objects (including fields) {\em appear}
classical to ``local" observers. It can, of course, not explain
why there are such local observers at all. The classical properties
are thereby defined by the {\em pointer basis} for the object,
which is distinguished by the interaction with the environment
and which is sufficiently stable in time. It is important to
emphasise that classical properties are {\em not} an a priori
attribute of objects, but only come into being through the
interaction with the environment.

Because decoherence acts, for macroscopic systems,
on an extremely short time scale, it appears to act discontinuously,
although in reality decoherence is a smooth process.
This is why ``events", ``particles", or ``quantum jumps" are being
observed. Only in the special arrangement of experiments,
where systems are used that lie at the border between microscopic
and macroscopic, can this smooth nature of decoherence be observed.

Since decoherence studies only employ the standard formalism of
quantum theory, all components characterising macroscopically
different situations are still present in the total quantum state
which includes system {\em and} environment, although they cannot be 
observed locally. Whether there is a
real dynamical ``collapse" of the total
state into one definite component or not (which would lead to an
Everett interpretation)
 is at present an undecided question.
Since this may not experimentally be decided in the near future,
it has been declared a ``matter of taste" (Zeh~1997).

Much of the discussion at this conference dealt with the question
of how a theory with ``definite events" can be obtained.
Since quantum theory without any collapse can immediately give
the {\em appearance} of definite events, it is important to understand
that such theories should possess additional features that make them
amenable to experimental test. For dynamical collapse models
such as the GRW-model or models invoking gravity (see Chap.~8 in
Giulini {\em et al.}~1996), the collapse
may be completely drowned by environmental decoherence, and would
thus not be testable, see in particular Bose, Jacobs, and Knight (1997)
for a discussion of the experimental situation.
 As long as no experimental hints about
testable additional features are available, such theories may be
considered as ``excess baggage", because quantum theory itself
can already explain everything that is observed. The price to pay,
however, is a somewhat weird concept of reality that includes for
the total quantum state all these macroscopically different
components.

The most important feature of decoherence besides its ubiquity
is its {\em irreversible} nature. Due to the interaction with the
environment, the quantum mechanical entanglement {\em increases}
with time. Therefore, the local entropy for subsystems increases, too,
since information residing in correlations is locally unobservable.
A natural prerequisite for any such irreversible behaviour,
most pronounced in the Second Law of thermodynamics, is a special
initial condition of very low entropy. Penrose has convincingly
demonstrated that this is due to the extremely special nature of the
big bang. Can this peculiarity be explained in any satisfactory way?
Convincing arguments have been put forward that this can only be
achieved within a quantum theory of gravity (Zeh~1992).
Since this discussion lies outside the scope of this contribution,
it will not be described here.

What is the ``Quantum Future" of decoherence? 
Two important issues play, in our opinion, a crucial role.
First, experimental tests should be extended to various situations
where a detailed comparison with theoretical calculations can be made.
This would considerably improve the confidence in the impact of
the decoherence process. It would also be important to study
potential situations where collapse models and decoherence would lead
to different results. This could lead to the falsification of
certain models. An interesting experimental situation
is also concerned with the construction of quantum computers
where decoherence plays the major negative role.
Second, theoretical calculations of concrete decoherence processes
should be extended and refined, in particular in {\em field} theoretical
situations. This could lead to a more profound understanding
of the superselection rules frequently used in these
circumstances.

\vskip 3mm
{\bf Acknowledgements}. C.K. thanks the organisers of the
Tenth Max Born Symposium, Philippe Blanchard and Arkadiusz Jadczyk,
for inviting him to this interesting and stimulating meeting.
%
%
%


\begin{thebibliography}
%
\bibitem{}{}{}
Bose, S., Jacobs, K., Knight, P.L. (1997):
A scheme to probe the decoherence of a macroscopic object.
Report quant-ph/9712017
%
\bibitem{}{}{}
Brune, M., Hagley, E., Dreyer, J., Ma\^{\i}tre, X., Maali, A.,
Wunderlich, C., Raimond, J.M., Haroche, S. (1996):
Observing the Progressive Decoherence of the ``Meter" in a
Quantum Measurement. Phys. Rev. Lett. {\bf 77}, 4887--4890
%
\bibitem{}{}{}
Caldeira, A.O., Leggett, A.J. (1983): Path integral approach
to quantum Brownian motion. Physica {\bf 121A}, 587--616
%
\bibitem{}{}{}
Demers, J.-G., Kiefer, C. (1996): Decoherence of black holes
by Hawking radiation. Phys. Rev. D {\bf 53}, 7050--7061
%
\bibitem{}{}{}
 Giulini, D., Joos, E., Kiefer, C., Kupsch, J.,
Stamatescu, I.-O., Zeh, H.D. (1996): {\it Decoherence and the
Appearance of a Classical World in Quantum Theory} (Springer, Berlin).
%
\bibitem{}{}{}
Habib, S., Kluger, Y., Mottola, E., Paz, J.P. (1996):
Dissipation and decoherence in mean field theory.
Phys. Rev. Lett. {\bf 76}, 4660--4663
%
\bibitem{}{}{}
Heisenberg, W. (1958): {\it Die physikalischen Prinzipien der
Quantentheorie}. (Bibliographisches Institut, Mannheim)
%
\bibitem{}{}{}
Jammer, M. (1974): {\it The Philosophy of Quantum
    Mechanics} (Wiley, New York)
%
\bibitem{}{}{}
Joos, E. (1984): Continuous measurement: Watchdog effect versus
golden rule. Phys. Rev. D {\bf 29}, 1626--1633
%
\bibitem{}{}{}
Joos, E. (1986): Why do we observe a classical spacetime?
Phys. Lett. A {\bf 116}, 6--8
%
\bibitem{}{}{}
Joos, E., Zeh, H.D. (1985): The emergence of classical properties
through interaction with the environment.
Z. Phys. B {\bf 59}, 223--243
%
\bibitem{}{}{} 
Kiefer, C. (1987): Continuous measurement of mini-superspace variables
by higher multipoles. Class. Quantum Grav. {\bf 4}, 1369--1382
%
\bibitem{}{}{}
Kiefer, C. (1992): Decoherence in quantum electrodynamics
and quantum cosmology. Phys. Rev. D {\bf 46}, 1658--1670
%
\bibitem{}{}{}
Kiefer, C., Polarski, P., Starobinsky, A.A. (1998): 
Quantum-to-classical transition for fluctuations in the early universe.
Submitted to Int. Journ. Mod. Phys. D [Report gr-qc/9802003]
%
\bibitem{}{}{}
K\"ubler, O., Zeh, H.D. (1973): Dynamics of quantum correlations.
Ann. Phys. (N.Y.) {\bf 76}, 405--418
%
\bibitem{}{}{}
Landau, L. (1927): Das D\"ampfungsproblem in der Wellenmechanik.
Z. Phys. {\bf 45}, 430--441
%
\bibitem{}{}{}
Mott, N.F. (1929): The wave mechanics of $\alpha$-ray tracks.
Proc. R. Soc. Lond. A {\bf 126}, 79--84
%
\bibitem{}{}{}
Omn\`{e}s, R. (1997): General theory of the decoherence effect
in quantum mechanics. Phys. Rev. A {\bf 56}, 3383--3394
%
\bibitem{}{}{}
Shaisultanov, R.Z. (1995a): Backreaction in scalar QED,
Langevin equation and decoherence functional. Report hep-th/9509154
%
\bibitem{}{}{}
Shaisultanov, R.Z. (1995b): Backreaction in spinor QED and
decoherence functional. Report hep-th/9512144
%
\bibitem{}{}{}
Zeh, H.D. (1970): On the interpretation of measurement in quantum
theory. Found. Phys. {\bf 1}, 69--76
%
\bibitem{}{}{}
Zeh, H.D. (1986): Emergence of classical time from a universal
wave function. Phys. Lett. A {\bf 116}, 9--12
%
\bibitem{}{}{}
Zeh, H.D. (1992): {\it The physical basis of the direction of time}
                  (Springer, Berlin)
%
\bibitem{}{}{}
Zeh, H.D. (1997): What is achieved by decoherence? In {\it
New Developments on Fundamental Problems in Quantum Physics}, edited by
M.~Ferrer and A.~van der Merwe (Kluwer Academic, Dordrecht)
[Report quant-ph/9610014]
%
\bibitem{}{}{}
Zurek, W.H. (1981): Pointer basis of quantum apparatus: Into what
mixture does the wave packet collapse? Phys. Rev. D {\bf 24},
1516--1525
%
\bibitem{}{}{}
 Zurek, W.H. (1991): Decoherence and the Transition
     from Quantum to Classical. Physics Today {\bf 44} (Oct.),
     36--44; see also the discussion in Physics Today (letters)
     {\bf 46} (April), 13
 %
\end{thebibliography}
\end{document}